\documentclass[runningheads]{svmult}

\usepackage{makeidx}   
\usepackage{graphicx}  
\usepackage{subeqnar}  
\usepackage{multicol}  
\usepackage{physprbb}  
\makeindex             

\begin{document}
\title*{Quantum Dots Attached to Ferromagnetic Leads: Exchange Field, Spin
Precession, and Kondo Effect}
\toctitle{Quantum Dots Attached to Ferromagnetic Leads: Exchange Field, Spin
Precession, and Kondo Effect}
%
%
\titlerunning{Quantum Dots Attached to Ferromagnetic Leads}
%
\author{J\"urgen K\"onig\inst{1}
 \and Jan Martinek\inst{2,3,4}
 \and J\'ozef Barna\'s\inst{4,5}
 \and Gerd Sch\"on\inst{2}}
%
%
%
\institute{Institut f\"ur Theoretische Physik III, Ruhr-Universit\"at Bochum, 44780 Bochum, Germany
\and
Institut f\"ur Theoretische Festk\"orperphysik, Universit\"at Karlsruhe, 76128 Karlsruhe, Germany
\and
 Institute for Materials Research, Tohoku University, Sendai
980-8577, Japan
 \and
Institute of Molecular Physics, Polish Academy of Sciences, 60-179
Pozna\'n, Poland
 \and
Department of Physics, Adam Mickiewicz University, 61-614
Pozna\'n,
 Poland}

\maketitle              

\begin{abstract}
Spintronics devices rely on spin-dependent transport behavior
evoked by the presence of spin-polarized electrons.
Transport through nanostructures, on the other hand, is dominated by strong 
Coulomb interaction. 
We study a model system in the intersection of both fields, a quantum dot 
attached to ferromagnetic leads. 
The combination of spin-polarization in the leads and strong Coulomb 
interaction in the quantum dot gives rise to an exchange field acting on
electron spins in the dot. 
Depending on the parameter regime, this exchange field is visible in the 
transport either via a precession of an accumulated dot spin or via an induced
level splitting. 
We review the situation for various transport regimes, and discuss two of 
them in more detail.
\end{abstract}

\section{Introduction}

The study of spin-dependent tunneling through quantum dots resides
in the intersection of two active and attractive fields of
physics, namely spintronics \cite{wolf,loss,maekawa} and transport
through nanostructures \cite{averin,nato,schon}. 
Both the investigation of spin-dependent electron transport on the one hand 
and the study of strong Coulomb interaction effects in transport through
nanostructures on the other hand define by now well-established
research areas. 
The combination of both concepts within one system is, however, a very new 
field which is still in its early stages. 
Its attractiveness originates from the rich physics expected from the 
combination of two different paradigms. 
A suitable model system for a basic study of the interplay of spin-dependent 
transport due to spin polarization in ferromagnetic electrodes and Coulomb 
charging effects in nanostructures is provided by a quantum dot attached to 
ferromagnetic leads.

\subsection{Some Concepts of Spintronics}

The field of spin- or magnetoelectronics \cite{wolf,loss,maekawa} has 
attracted much interest, for both its beautiful fundamental physics and its 
potential applications. 
A famous example, which has already proven technological relevance, is the 
spin valve based on either the giant magnetoresistance effect (GMR) in
magnetic multilayers or the tunnel magnetoresistance (TMR) in magnetic tunnel 
junctions.
In both cases, the transport properties depend on the relative magnetization
orientation of the involved magnetic layers or leads, an information conveyed 
by the spin polarization of the transported electrons. 
In the case of a single magnetic tunnel junction, the tunneling current is 
maximal for parallel alignment of the leads' magnetization orientations, while 
it is minimal for antiparallel alignment. 
This can be easily understood within a non-interacting-electron picture, as 
proposed by Julli{\`e}re \cite{julliere}: the tunnel current of electrons with 
given spin direction is proportional to the product of the corresponding
spin-dependent densities of states in the source and drain electrode, which 
leads to a reduction of transport in the case of antiparallel alignment.

\begin{figure}
\centerline{\includegraphics[width=10.cm]{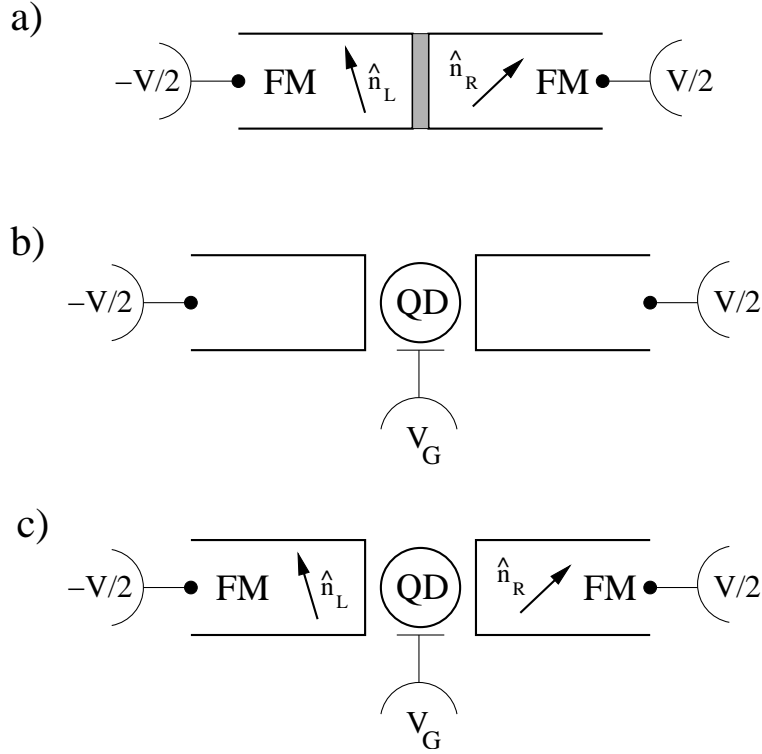}} \caption{
  a) Spin valve: a single tunnel junction between two ferromagnets (FM) with
  magnetization orientations ${\bf \hat n}_{\rm L}$ and
  ${\bf \hat n}_{\rm R}$, respectively.
  b) Quantum dot.
  c) Quantum-dot spin valve: a quantum dot is connected to two ferromagnetic
  leads (FM)}
\label{fig1}
\end{figure}

This concept has been extended \cite{slonczewski} to describe also
noncollinear arrangements, as depicted in Fig.~\ref{fig1}(a), where the 
magnetization directions of the leads enclose an arbitrary angle $\phi$. 
In this situation, the $\phi$-dependent part of the tunneling current is 
proportional to the overlap of the spinor part of the majority-spin wave 
functions in the source and drain electrode, i.e. proportional to $\cos \phi$, 
as it has been experimentally confirmed recently \cite{angular}.

In heterostructures that consist of a nonmagnetic metal sandwiched
by ferromagnetic electrodes, the concept of spin accumulation
becomes important. 
Once the spin diffusion length is larger than the size of the nonmagnetic 
region, the information about the relative orientation of the leads' 
magnetization is mediated through the middle part. 
In the antiparallel configuration an applied bias voltage leads to a pile up 
of spin in the nonmagnetic metal, since electrons with one type of spin
(say spin up) are preferentially injected from the source electrode, while 
electrons with the other type of spin (spin down) are pulled out from the 
drain electrode. 
This piling up of spin splits the chemical potentials for spin-up and spin-down
electrons in the normal metal such that electrical transport through the 
whole device is reduced.

As spin is a vector quantity, transport through a
ferromagnetic-nonmagnetic-ferromagnetic heterostructure can be tuned by 
manipulating the direction of the spins in the middle part. 
The prototype for such a concept is the spin field-effect transistor proposed 
by Datta and Das \cite{datta}. 
Spin-polarized electrons are injected from a ferromagnetic metal into a 
ballistic conducting channel provided by a two-dimensional electron gas in a
semiconductor heterostructure. 
Due to the Rashba effect, the electrons in the semiconductor experience a 
spin-orbit coupling, whose strength can be tuned by a gate voltage. 
This spin-orbit coupling leads to a rotation of the spins in the conducting
channel as they move along towards the drain electrode. 
The total transmission through the device, then, depends on the relative
orientation of the rotated spins and the magnetization direction of the drain 
electrode.

\subsection{Transport Through Nanostructures}

Tunneling transport through nanostructures, such as semiconductor
quantum dots (Fig.~\ref{fig1}(b)) or small metallic islands, is
strongly affected by Coulomb interaction, and a non-interacting
electron picture is no longer applicable \cite{averin,nato,schon}.
Coulomb-blockade phenomena arise at low temperature, such that the 
corresponding energy scale is smaller than the charging energy, the energy 
scale for adding or removing one electron from the dot or island. 
Small quantum dots with a size of the order of the Fermi wavelength have a 
discrete level spectrum. 
If the level spacing is large enough, transport through single levels is 
possible. 
This situation defines a simplest but very generic model, the Anderson-impurity
model, for studying Coulomb interaction in nanostructures.

When the level is occupied with one electron since double occupancy is 
prohibited by charging energy, the dot possesses a local spin. 
At low temperature and large dot-lead tunnel-coupling strength, a ground state 
with complex manybody correlations forms, which manifests itself in the 
so-called Kondo effect \cite{hewson-book}. 
The local spin is screened by the spins of the conduction electrons in the 
leads, and accompanied with this, electrical transport through the quantum dot 
is strongly enhanced.

\subsection{Quantum-Dot Spin Valves}

The scheme of a quantum-dot spin valve, a quantum dot attached to 
ferromagnetic leads, is illustrated in Fig.~\ref{fig1}(c).
Successful fabrication of either quantum-dot systems or magnetic 
heterostructures has been achieved by a large number of experimental groups. 
To attach ferromagnetic electrodes to quantum dots, though, is quite a 
challenging task, and only very recently first results have been reported.

Let us start with metallic single-electron devices. 
Both spin-dependent tunneling and Coulomb blockade has been found in
magnetic tunnel junction with embedded Co clusters \cite{schelp}.
All-ferromagnetic metallic single-electron transistors have been
manufactured, using either single-island \cite{brueckl,ono}, or multi-island
structures \cite{mitani1,mitani2}. 
Magnetoresistance of single-electron transistors with a normal metallic island 
in a cobalt-aluminum-cobalt structure has been measured \cite{chen}. 
In all these examples, the level spectrum on the island is continuous, and 
many levels are involved in transport.

Our focus, however, is on single-level quantum dots. 
The difficulty lies in the incompatibility of the usual materials showing 
ferromagnetism (metals) and those usually forming quantum dots 
(semiconductors). 
There are different strategies to overcome this problem. 
One possibility is the use of ferromagnetic semiconductors (Ga,Mn)As as lead 
electrodes coupled to, e.g., self-assembled InAs quantum dots \cite{chye}. 
A very promising approach is to contact an ultrasmall aluminum nanoparticle, 
which serves as a quantum dot, to ferromagnetic metallic electrodes. 
In this way, quantum dots with one magnetic (nickel or cobalt) and one 
nonmagnetic (aluminum) electrodes have been fabricated \cite{ralph1}. 
Another important system is a magnetic impurity inside
the tunneling barrier of ferromagnetic tunnel junction \cite{impurity}.
An alternative route is to use carbon nanotubes as quantum dots and to place 
them on metallic contacts.
Coulomb-blockade phenomena and even the Kondo effect has been observed in such 
systems \cite{nygard1,buitelaar}. 
Spin-dependent transport through carbon nanotubes attached to ferromagnetic 
electrodes has been investigated in \cite{cnt,nygard2}.
A more challenging scheme is a ferromagnetic single-molecule transistor 
\cite{ralph2}, where a single molecule is attached to ferromagnetic electrodes.
To some extent, there is also a relation between the quantum-dot spin valve 
and a  single magnetic-atom spin on a scanning tunneling microscope tip. 
For the latter, precession of the single spin in an external magnetic field has
been detected in the power spectrum of the tunneling current 
\cite{precession_e}.

This progress on the experimental side has stimulated a number of theoretical
activities \cite{metal-theory,QD-theory,martinek3,koenig,braun,weymann,sergueev,zhang,bulka,lopez,martinek1,martinek2,choi,martinek4} on spin-dependent 
transport through either metallic single-electron transistors or quantum dots.

The motivation for studying quantum-dot spin valves can be formulated from 
two different perspectives, depending on from which side one starts to 
approach the problem. 
Coming from the spintronics side, one may ask how the concepts introduced 
there, such as spin accumulation and spin manipulation, manifest themselves in 
quantum dots, and how the presence of strong Coulomb interaction gives rise to 
qualitatively new behavior as compared to non-interacting electron systems. 
On the other hand, when starting from Coulomb-interaction effects in quantum 
dots, one may ask how the spin-polarization of the leads changes the picture. 
As mentioned above, the screening of a local spin on the quantum dot by the 
lead-electron spins is crucial for the Kondo effect to develop. 
This screening behavior is affected by spin asymmetry introduced due to 
a finite spin polarization of the lead.
In this case, it is a priori not clear whether a Kondo-correlated state
can still form or not.

To comprise all this in a single question, we ask whether the combination of 
strong Coulomb interaction and finite spin-polarization gives rise to 
qualitatively new phenomena that are absent for either non-interacting or 
unpolarized electrons. 
The answer is: yes, it does.
We predict that single electrons on the quantum dot experience an exchange 
field, which effectively acts like a local magnetic field.
The main goal of this paper is to illustrate the origin of this exchange field,
its properties, and its implications on transport.
Of course, the latter depends on the considered transport regime, and the
observable consequences can be quite different.
In the present paper, we concentrate on two particular regimes, the case of 
weak dot-lead coupling but noncollinear magnetization directions and the case 
of very strong coupling but collinear configuration. 
Other limits will only be commented on shortly, as for these cases work is 
still in progress and will be presented elsewhere.

\section{The Model}

We consider a small quantum dot with one energy level $\epsilon$ participating 
in transport. 
The dot is coupled to ferromagnetic leads, see Fig.~\ref{fig1}(c).
The left and right lead are magnetized along $\hat \mathbf{n}_\mathrm{L}$ 
and $\hat \mathbf{n}_\mathrm{R}$, respectively.
The total Hamiltonian is

\begin{equation}
  H = H_\mathrm{dot} +
  H_\mathrm{L} + H_\mathrm{R} + H_\mathrm{T,L} + H_\mathrm{T,R} \, .
\label{hamiltonian}
\end{equation}

The first part, $ H_\mathrm{dot} = \epsilon \sum_\sigma 
c^\dagger_\sigma c_\sigma + U n_\uparrow n_\downarrow$, describes the dot 
energy level plus the charging energy $U$ for double occupation. 
In the presence of an external magnetic field, the energy level experiences
a Zeeman splitting, i.e., becomes spin-dependent.
The leads are modeled by 
$H_r = \sum_{k\sigma} \epsilon_{k \sigma} a^\dagger_{rk\sigma} a_{rk\sigma}$ 
with $r = \mathrm{L,R}$. 
In the spirit of a Stoner model of ferromagnetism \cite{yosida}, there is a 
strong spin asymmetry in the density of states $\rho_{r\sigma}(\omega)$ for
majority ($\sigma=+$) and minority ($\sigma=-$) spins. 
Throughout all of our calculations presented here, we approximate the density
of states to be energy independent, 
$\rho_{r\sigma}(\omega)= \rho_{r\sigma}$. 
Real ferromagnets will have a structured density of states \cite{papa}. 
This fact, however, will only modify details of the results and not the main 
physical picture.
The ratio $p = (\rho_{r+}-\rho_{r-})/(\rho_{r+}+\rho_{r-})$ characterizes the 
degree of spin polarization in the leads. 
For simplicity, we assume here $\rho_{\rm{L}+}=\rho_{\rm{R}+}\equiv
\rho_+$ and $\rho_{\rm{L}-}=\rho_{\rm{R}-}\equiv \rho_-$.
Nonmagnetic leads are described by $p=0$, and $p=1$ represents half metallic 
leads, which accommodate majority spins only. 
We emphasize that the magnetization directions of leads can differ from each 
other, enclosing an angle $\phi$.

Tunneling between leads and dot is described by the standard 
tunneling Hamiltonian.
For the left tunnel barrier we get

\begin{equation}
  H_{\mathrm{T},\mathrm{L}} = t \sum_{k\sigma=\pm} \left( 
  a^\dagger_{\mathrm{L}k\sigma}
  c_\sigma + h.c. \right) \, ,
\label{tunnel1}
\end{equation}

where $c_\pm$ are the Fermi operators for an electron on the quantum dot with 
spin along $\pm \hat{\vec{n}}_r$.
For the right barrier, an analogous expression holds.
As $\hat{\vec{n}}_{\rm L}$ may differ from $\hat{\vec{n}}_{\rm R}$, an ambiguity
arises in the definition of $c_\pm$.
This is no problem for collinear, i.e., parallel or antiparallel, 
configuration of the leads.
In this case, $\hat{\vec{n}}_\mathrm{L} = \pm \hat{\vec{n}}_\mathrm{R}$ provides 
a natural quantization axis for the dot spin.

\begin{figure}
\centerline{\includegraphics[width=8.cm]{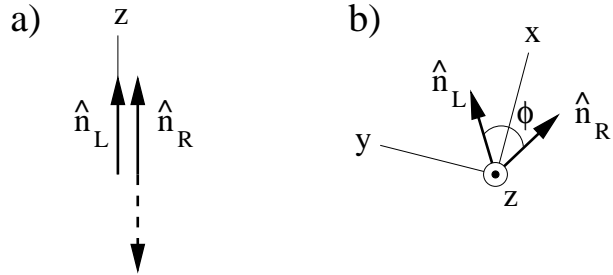}}
\caption{
  Choice of the used coordinate system:
  a) For collinear configuration of the leads' magnetization, i.e., parallel
  (solid arrow for $\hat{\vec{n}}_{\rm R}$) or antiparallel (dashed arrow),
  the $z$-axis is along $\hat{\vec{n}}_{\rm L}$.
  In this case, we use the tunneling Hamiltonian in the form of
  (\ref{tunnel1}).
  b) For noncollinear arrangements, the $z$-axis is perpendicular to both
  $\hat{\vec{n}}_{\rm L}$ and $\hat{\vec{n}}_{\rm R}$.
  Here, the tunneling Hamiltonian in the form of (\ref{tunnel2}) is used.
  The quantum-dot spin is always quantized along the $z$-axis}
\label{fig2}
\end{figure}

For noncollinear leads, however, the form (\ref{tunnel1}) of the 
tunnel Hamiltonian is no longer useful.
To describe the scenario properly, we find it convenient to
quantize the dot spin neither along $\hat{\vec{n}}_\mathrm{L}$ nor
$\hat{\vec{n}}_\mathrm{R}$, but along the axis perpendicular to both
$\hat{\vec{n}}_{\rm L}$ and $\hat{\vec{n}}_{\rm R}$. 
To be explicit, we choose the coordinate system defined by
$\hat{\vec{e}}_x = (\hat{\vec{n}}_\mathrm{L}+ \hat{\vec{n}}_\mathrm{R}) /
|\hat{\vec{n}}_\mathrm{L}+\hat{\vec{n}}_\mathrm{R}|$, 
$\hat{\vec{e}}_y = (\hat{\vec{n}}_\mathrm{L}- \hat{\vec{n}}_\mathrm{R}) /
|\hat{\vec{n}}_\mathrm{L}-\hat{\vec{n}}_\mathrm{R}|$, and 
$\hat{\vec{e}}_z = (\hat{\vec{n}}_\mathrm{R} \times 
\hat{\vec{n}}_\mathrm{L})/ |\hat{\vec{n}}_\mathrm{R} 
\times \hat{\vec{n}}_\mathrm{L}|$, and quantize the dot spin along the
$z$-direction, see Fig.~\ref{fig2}. 
The tunnel Hamiltonian, then, becomes

\begin{equation}
  H_\mathrm{T,L} = {t\over \sqrt{2}} \sum_k
   (a^\dagger_{\mathrm{L}k+}, a^\dagger_{\mathrm{L}k-})
  \left( \begin{array}{cc}
    e^{i\phi/4} & e^{-i\phi/4} \\
    e^{i\phi/4} & - e^{-i\phi/4}
  \end{array} \right)
  \left( \begin{array}{c} c_\uparrow \\ c_\downarrow \end{array} \right)
  + h.c.
  \, ,
\label{tunnel2}
\end{equation}

and $H_\mathrm{T,R}$ is the same but with $\mathrm{L} \rightarrow \mathrm{R}$ 
and $\phi \rightarrow - \phi$. 
The special choice of the coordinate system implies that both up and down 
spins of the dot are equally-strongly coupled to the majority and minority 
spins of the leads. 
There, are, however, phase factors $e^{\pm i\phi/4}$ are involved, similar to 
multiply-connected quantum-dot systems dubbed Aharonov-Bohm 
interferometers \cite{AB}. 
The two spin directions $\uparrow$ and $\downarrow$ in the dot correspond to 
the quantum dots placed in the two arms of the Aharonov-Bohm interferometer, 
and the angle $\phi$ plays the role of the Aharonov-Bohm phase, which measures 
the total magnetic flux enclosed by the arms of the interferometer in units of 
the flux quantum. 
We note, however, that our model translates to a very special kind of 
Aharonov-Bohm interferometer: the dot in each interferometer arm accommodates 
only a single level instead of a doubly-degenerate one, and Coulomb 
interaction occurs between the two dots, instead of within each of them.

The two different choices we use for the collinear and noncollinear 
configuration, in which we use either (\ref{tunnel1}) or 
(\ref{tunnel2}), respectively, are illustrated in Fig.~\ref{fig2}. 
In both cases, the tunnel coupling leads to a finite width of the dot level.
Its energy scale is given by
$\Gamma = \sum_r \Gamma_r$ with $\Gamma_r = \pi |t|^2 \sum_{\sigma=\pm}
\rho_{\sigma}$ \cite{com_models}.

\section{Exchange Field}

As pointed out in the introduction, the qualitative new physics introduced 
by the combination of spin-polarized leads and strong Coulomb interaction in 
the dot, is the existence of an exchange field acting on electron spins in the
dot.
This exchange field is intrinsically present in the model described by 
the Hamiltonian (\ref{hamiltonian}) together with the spin-dependent 
density of states.
It is, therefore, automatically contained in any consistent treatment of the 
model for a given transport regime, as we will see in the subsequent sections. 
Nothing has to be added by hand.
Nevertheless, we find it instructive to derive an explicit analytic expression 
by making use of the following heuristic procedure.

Each of the two leads will contribute to the exchange field separately.
To keep the discussion transparent, we consider the effect of one lead only. 
The total exchange field is, then, just the sum over both leads. 
The first step is to derive an effective Hamiltonian for the subspace of the 
total Hilbert space in which the quantum dot is singly occupied. 
This is the regime of interest, as far as the exchange field in concerned, 
since both an empty and a doubly-occupied dot have zero total spin, and an
exchange field would be noneffective. 
By taking into account virtual excitations to an empty or doubly-occupied dot 
within lowest-order perturbation theory in the tunnel coupling, in analogy to 
the Schrieffer-Wolff transformation \cite{hewson-book} employed in the context 
of Kondo physics for magnetic impurities in nonmagnetic metals, we arrive at 
an effective spin model for the dot spin operators $S^\pm$ and $S^z$
(quantized along the magnetization direction of the considered lead),

\begin{eqnarray}
  H_{\rm spin} &=& S^+  |t|^2 \sum_{kq}
  \left( \frac{1}{U+\epsilon - \epsilon_q}
  + \frac{1}{\epsilon_k - \epsilon} \right)
  a^\dagger_{rk\downarrow}a_{rq\uparrow}
\nonumber \\
  &&
  + S^- |t|^2 \sum_{kq}
  \left( \frac{1}{U+\epsilon - \epsilon_k}
  + \frac{1}{\epsilon_q - \epsilon} \right)  
  a^\dagger_{rq\uparrow}a_{rk\downarrow}
\nonumber \\
  &&
  + S^z |t|^2 \left( \sum_{q q'} \frac{1}{U+\epsilon - \epsilon_{q'}}
  a^\dagger_{rq\uparrow} a_{rq'\uparrow}
  - \sum_{k k'} \frac{1}{U+\epsilon - \epsilon_{k'}}
  a^\dagger_{rk\downarrow}a_{rk'\downarrow}
  \right)
\nonumber \\
  &&
  - S^z |t|^2 \left( \sum_{q q'} \frac{1}{\epsilon_{q} - \epsilon}
  a_{rq'\uparrow} a^\dagger_{rq\uparrow}
  - \sum_{k k'} \frac{1}{\epsilon_{k} - \epsilon}
  a_{rk'\downarrow}a^\dagger_{rk\downarrow}
  \right)
\, .
\label{spinham}
\end{eqnarray}

Note, that the information about the different densities of states for up- and
down-spins is included in the summation over $q, q'$  (used for spin-up
electrons) and $k, k'$ (used for spin down), respectively. 
In addition, there is a term describing potential scattering, but this does 
not contribute to the exchange field we are aiming at.

\begin{figure}
\centerline{\includegraphics[width=10.cm]{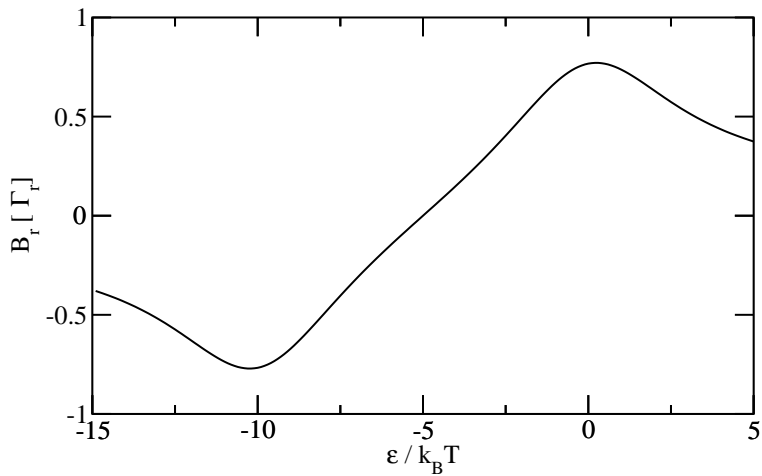}}
\caption{
  The exchange field as a function of the level position $\epsilon$ for
  $U/k_{\rm B}T = 10$ and $p=1$}
\label{fig3}
\end{figure}

In a second step we employ in (\ref{spinham}) a mean-field approximation
for the lead-electron states, making use of
$\langle a^\dagger_{rk\sigma}a_{rk'\sigma'} \rangle = f_r(\epsilon_{k\sigma})
\delta_{kk'} \delta_{\sigma\sigma'}$
and
$\langle a_{rk\sigma}a^\dagger_{rk'\sigma'} \rangle = 
[1-f_r(\epsilon_{k\sigma})]\delta_{kk'} \delta_{\sigma\sigma'}$,
where $f_r(\omega)$ is the Fermi function
of lead $r$.
The terms proportional to $S^\pm$ drop out.
The resulting effective Hamiltonian, then, reads $H_{\rm eff} =- S^z B_r$
with the exchange field (for simplicity we include the gyromagnetic factor in 
the definition)

\begin{eqnarray}
  B_r &=& \int' d \omega (\rho_+ - \rho_-) |t|^2 \left(
 \frac{1-f_r(\omega)}{\omega - \epsilon}
  + \frac{f_r(\omega)}{\omega - \epsilon - U} \right)
\label{heff}
\\
  &=& - {p \Gamma_r \over \pi} \mathrm{Re} \left[
    \Psi\left({1\over 2}+i{\beta(\epsilon-\mu_r)\over 2\pi}\right)
    -\Psi\left({1\over 2}+i{\beta(\epsilon+U-\mu_r)\over 2\pi}\right)
\right] \, ,
\label{exchange}
\end{eqnarray}

where $\Psi(x)$ denotes the digamma function, $\mu_r$ is the electrochemical 
potential of lead $r$, and the prime at the integral sign in (\ref{heff}) 
symbolizes Cauchy's principal value. 
For illustration, we plot the exchange field as a function of the level 
position in Fig.~\ref{fig3}.

\vspace*{.2cm}

From the explicit form (\ref{exchange}) of the exchange field we derive
the following properties:
\begin{itemize}
\item[(i)] It vanishes in the case of a non-interacting quantum dot, $U=0$.
\item[(ii)] The exchange field is proportional to the degree of
spin-polarization $p$ in the lead.
This means that both strong Coulomb interaction and finite spin-polarization
are required to generate the exchange field.
\item[(iii)] It depends on the tunnel coupling strength $\Gamma$.
In the treatment lined out above, $\Gamma$ enters linearly as a global
prefactor.
\item[(iv)] The magnitude and even the sign of the exchange field depends on
the level position $\epsilon$.
In particular, there is a value of $\epsilon$ at which the exchange field
vanishes (in our model with flat density of states this happens at
$\epsilon - \mu_r = U/2$, i.e., when the total system is particle-hole
symmetric).
\end{itemize}

Furthermore, we notice from (\ref{heff}) that not only electronic states 
around the Fermi energy of the lead are involved.
Instead, it is rather the full band that matters. 
This means, that a precise simulation of realistic materials requires a
knowledge of the detailed density of states, to be inserted in the integral 
in (\ref{heff}). 
This will modify the details of the exchange field such as its precise 
dependence on the level position $\epsilon$.

\section{Transport Regimes}

After introducing the notion of the exchange field, the question of how it
affects the transport behavior arises immediately.
The answer to this question depends on the transport regime under
consideration.
In particular, we will identify two mechanisms by which the exchange field
enters.
One scenario is the generation of a level splitting between up and down spins
in the quantum dot, with the level splitting given by the exchange field
(\ref{exchange}).
But this is not the only possibility.
Even in situations, in which the generated level splitting is negligible,
the exchange field can affect the dot state and, thus, the transport behavior
by rotating an accumulated spin on the dot, which can pile up there in
non-equilibrium due to an applied bias voltage.
A complete picture of the various different transport regimes goes beyond the
scope of the present paper.
Instead, we will concentrate on two specific limits, namely weak dot-lead
coupling but noncollinear magnetization in linear response, and strong
coupling but collinear configuration of the leads.
For some other regimes, that are currently under investigation, we will only
give some short comments and refer the reader to forthcoming publications.

In the limit of weak dot-lead coupling, $\Gamma \ll k_{\rm B}T$,
referred to as sequential-tunneling regime, transport is dominated
by processes of first order in $\Gamma$ (unless both $\epsilon$ and 
$\epsilon+U$ are shifted into the Coulomb-blockade region). 
First-order transport probes the state of the quantum dot to zeroth order 
(since the tunneling between dot and leads necessary for transport already 
trivially involves a factor $\Gamma$). 
Therefore, the level splitting generated by the exchange field cannot be 
probed by first-order transport. 
Nevertheless, the exchange field plays a role via the second of the above 
mentioned mechanisms. 
Once a finite spin is accumulated on the quantum dot, with a direction
noncollinear to the exchange field, the latter will induce a precession of the 
accumulated spin. 
For this to happen, a noncollinear configuration of the leads' magnetic 
moments is required, as otherwise accumulated spin, if any, and exchange
field are pointing in the same direction.

In the Coulomb-blockade regime, sequential tunneling is exponentially 
suppressed, and transport is dominated by cotunneling, which
are second-order processes.
But also on resonance, second-order corrections become important for
intermediate coupling strengths, $\Gamma \sim k_{\rm B}T$.
Second-order transport is affected by the generated level splitting, and the
exchange field plays a role even for a collinear arrangement of the leads'
magnetization.

A very dramatic signature of the level splitting generated by the
exchange field is predicted for the limit of low temperature and
large coupling strength, $k_{\rm B} T \le k_{\rm B} T_{\rm K} \ll
\Gamma$, for which the Kondo effect can appear ($T_{\rm K}$ is the
Kondo temperature). 
Since a finite level splitting, e.g., due to a Zeeman term induced by an 
external magnetic field, quickly destroys the Kondo effect, the exchange field 
has quite an important, at first glance destructive, consequence. 
As we will see below in more detail, however, by applying an 
appropriately-tuned external magnetic field one can compensate for the induces 
level splitting and, thus, recover the Kondo effect. 
For this discussion, we restrict ourselves to collinear configurations.

\subsection{First-Order Transport in Linear Response}

Here, we only present the major steps and main results.
Details of the calculations can be found in Refs.~\cite{koenig,braun}. 
The first step is to relate the linear conductance 
$G ^\mathrm{lin} = (\partial I / \partial V)\big|_{V=0}$ to the Green's 
functions of the dot. 
For first-order transport, we obtain

\begin{eqnarray}
   G^\mathrm{lin} &=&
        {e^2 \over h} \Gamma \int d \omega \, \left\{
        {\rm Im} \, G_{\downarrow\downarrow}^{\mathrm{ret}}(\omega) f' (\omega)
\right. \nonumber \\
        &&\left. \qquad \qquad \,\,\,\,\,
        + p\sin{\phi \over 2} \left[ f(\omega) \,
        {\partial G^>_{\downarrow\uparrow}(\omega) \over \partial (eV)}
        + \left[ 1 - f(\omega) \right]
        {\partial G^<_{\downarrow\uparrow}(\omega) \over \partial (eV)}
        \right] \right\} \, .
\label{Greens}
\end{eqnarray}

Here, $f(\omega)$ is the Fermi function, $G_{\sigma\sigma'}(\omega)$ are the 
Fourier transforms of the usual retarded, greater and lesser Green's functions.
Contributions involving the Green's functions $G_{\uparrow\uparrow}(\omega)$ 
and $G_{\uparrow\downarrow}(\omega)$ are accounted for in a prefactor 2. 
Since $\Gamma$ already appears explicitly in front of the integral, all 
Green's functions are to be taken to zeroth order in $\Gamma$. 
In this limit, we find $-(1/ \pi) \mathrm{Im} \, 
G^{\mathrm{ret}}_{\downarrow\downarrow} (\omega) = ( P^0_0 +
P^\downarrow_\downarrow ) \delta(\omega - \epsilon) +
(P^\uparrow_\uparrow + P^d_d ) \delta(\omega - \epsilon - U)$,
$G^>_{\downarrow\uparrow} (\omega) = 2\pi i P^\downarrow_\uparrow
\delta(\omega - \epsilon - U)$, and 
$G^<_{\downarrow\uparrow} (\omega) = 2\pi i P^\downarrow_\uparrow 
\delta(\omega - \epsilon)$, where
$P^\chi_{\chi'} = \langle |\chi'\rangle\langle \chi| \rangle$ are
elements of the stationary density matrix (to zeroth order in
$\Gamma$) of the quantum-dot subsystem, with $\chi,\chi' = 0$
(empty dot), $\uparrow, \downarrow$ (singly-occupied dot), and $d$
(doubly-occupied dot).

The main task is now to determine the density-matrix elements to
zeroth order in $\Gamma$. 
They contain the information about the average occupation and spin on the 
quantum dot. 
The diagonal matrix elements, $P^\chi_{\chi}$, are nothing but the 
probabilities to find the quantum dot in state $\chi$, i.e., the dot is empty 
with probability $P_0 \equiv P_0^0$, singly occupied with $P_1 \equiv
P_\uparrow^\uparrow + P_\downarrow^\downarrow$, and doubly occupied with 
$P_d \equiv P_d^d$. 
A finite spin can only emerge for single occupancy. 
The average spin $\hbar\vec{S}$ with $\vec{S} = (S_x,S_y,S_z)$ is related
to the matrix elements $P^\chi_{\chi'}$ via 
$S_x = \mathrm{Re} \,
P^\downarrow_\uparrow$, $S_y = \mathrm{Im} \,
P^\downarrow_\uparrow$, and $S_z = (1/2) (P^\uparrow_\uparrow -
P^\downarrow_\downarrow)$.
To obtain the density-matrix elements by using the real-time transport
theory developed in Ref.~\cite{diagrams}, we solve a kinetic equation
formulated in Liouville space.
The details are found in Refs.~\cite{koenig,braun}.

It is remarkable that on the r.h.s of (\ref{Greens}), derivatives of 
Green's function with respect to bias voltage $V$ appear. 
As a consequence, the linear conductance is not only determined by equilibrium 
properties of the quantum dot, but linear corrections in $V$ are involved as 
well. 
This is consistent with the observation that, in equilibrium, the density 
matrix is diagonal with the matrix elements determined by the Boltzmann 
factors, i.e., the average spin on the quantum dot vanishes at $V=0$ 
\cite{comment_2}. 
With applied bias voltage, though, a finite spin can accumulate. 
Therefore, to be sensitive to the relative magnetization direction of the 
leads, the linear conductance has to be connected to the differential spin 
accumulation $(d\vec{S}/dV)\big|_{V=0}$.

The results we find can be summarized as follows. 
At finite bias voltage, spin accumulated on the dot. 
Here, we only need its contribution linear in $V$ and find

\begin{equation}
  \frac{\partial|\vec{S}|}{\partial (eV)} \bigg|_{V=0} =
       {p P_1\over 4k_B T} \cos \alpha(\phi) \sin {\phi \over 2} \, ,
\end{equation}

where $P_1$ is the equilibrium probability for a singly occupied dot. 
The spin is lying in the $y$-$z$-plane enclosing an angle $\alpha$ with the 
$y$-axis, where
\begin{equation}
   \tan \alpha(\phi)
 = - {B \over \Gamma [1-f(\epsilon)+f(\epsilon+U)]} \cos{\phi \over 2} \, .
\label{angle}
\end{equation}

In the absence of an exchange field, the accumulated spin is oriented 
along $\hat{\vec{n}}_L - \hat{\vec{n}}_R$, i.e., it has a $y$-component 
only, $ \alpha = 0 $. 
The exchange field $B$, though, leads to a precession of the spin about the 
$x$-axis. 
The factor $1/\Gamma [1-f(\epsilon)+f(\epsilon+U)]$ in (\ref{angle}) can be
identified as the life time of the dot spin, limited by tunneling out of the 
dot electron or by tunneling in of a second electron with opposite spin. 
Since both this life time and the exchange field are of first order in 
$\Gamma$, the angle $\alpha$ acquires a finite value.

\begin{figure}[h]
\centerline{\includegraphics[width=10cm]{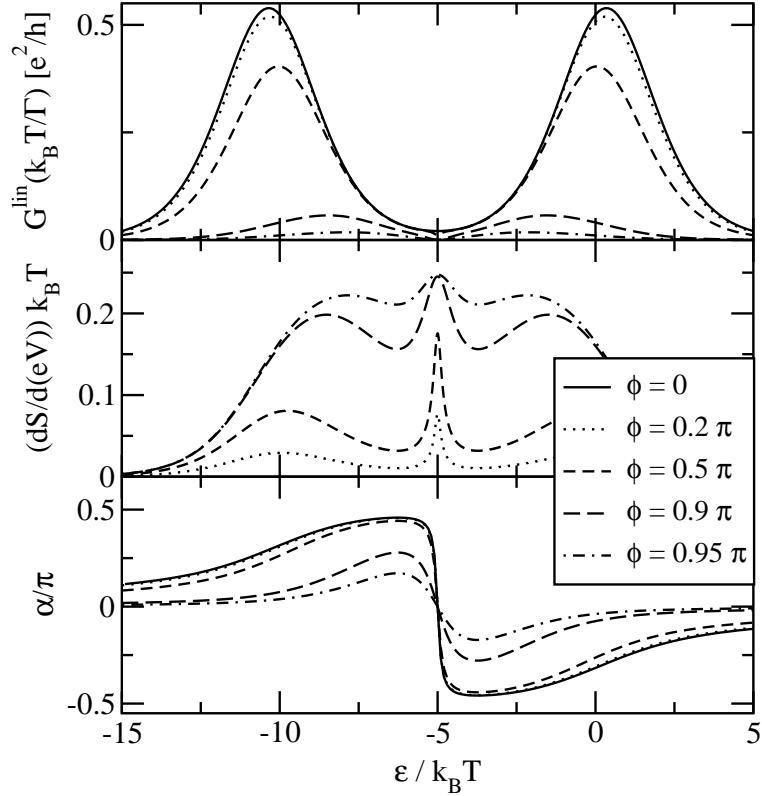}} 
\caption{
  Upper panel: Linear conductance (normalized by $\Gamma/k_BT$ and plotted
  in units of $e^2/h$) as a function of level position $\epsilon$
  for five different angles $\phi$. Middle panel: Derivative of
  accumulated spin $S$ with respect to bias voltage $V$ normalized
  by $k_BT$. Lower panel: angle $\alpha$ between the quantum-dot
  spin and the $y$-axis. In all panels we have chosen the charging
  energy $U/k_BT = 10$ and $p=1$ }
\label{fig4}
\end{figure}

The differential spin accumulation $dS/d(eV)$ in units of $k_BT$ is
illustrated in the middle panel of Fig.~\ref{fig4}. 
It is clear that single occupation of the dot is required for spin
accumulation, i.e., the plotted signal is high in the valley
between the two conductance peaks. 
The lower panel of Fig.~\ref{fig4} shows the evolution of the rotation angle
$\alpha$ as a function of the level energy $\epsilon$. 
This angle is large in the valley between the conductance peaks, getting close 
to $\pm \pi/2$. 
A special point is $\epsilon = -U/2$, at which, due to particle-hole symmetry, 
the exchange interaction vanishes. 
As a consequence, $\alpha$ shows a sharp transition from positive to
negative values, accompanied with a peak in the accumulated spin.

The linear conductance is given by

\begin{equation}
   G^\mathrm{lin} = G^\mathrm{lin, max}
   \left( 1 - p^2 \cos^2 \alpha(\phi) \sin^2{\phi \over 2} \right) \, .
\end{equation}

The conductance is maximal for parallel magnetization, $\phi=0$.
Its value is $G^\mathrm{lin, max} = (\pi e^2/ h) (\Gamma/ k_BT)
[1-f(\epsilon+U)]f(\epsilon) [1-f(\epsilon)+f(\epsilon+U)] /
[f(\epsilon)+1-f(\epsilon+U)]$. 
The upper panel of Fig.~\ref{fig4} depicts the linear conductance for five 
different values of the angle $\phi$. 
For parallel magnetization, $\phi=0$, there are two conductance peaks located 
near $\epsilon=0$ and $\epsilon = -U$, respectively. 
With increasing angle $\phi$, transport is more and more suppressed due to the 
spin-valve effect. 
However, this suppression is not uniform, as would be in the absence of the
exchange field. 
In contrast, the spin-valve effect is less pronounced in the valley between the
two peaks, where the rotation angle $\alpha$ is large. 
A large angle $\alpha$ reduces both the magnitude of the accumulated spin, as 
discussed above, and the relative angle to the magnetization of the drain 
electrode. 
Both enhance transport as compared to the situation without the exchange field.
As a consequence, the two conductance peaks move towards each other with 
increasing $\phi$.

\begin{figure}
\centerline{\includegraphics[width=10cm]{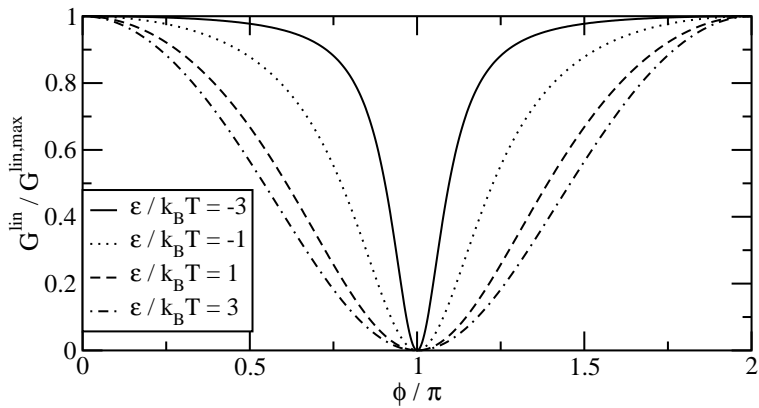}}
\caption{Normalized linear conductance as a function of $\phi$ for
  $U/k_BT = 10$, $p=1$, and four different values of the level position}
\label{fig5}
\end{figure}

Another way to illustrate the influence of the exchange field is to plot the 
$\phi$-dependence of the linear conductance, see Fig.~\ref{fig5}. 
For values of the level position $\epsilon$ at which the rotation angle 
$\alpha$ is small, $\epsilon/k_BT = 3$ and $1$, the $\phi$-dependence of the 
conductance is almost harmonic, as it is for single magnetic tunnel junction. 
For $\epsilon/k_BT = -1$ and $-3$, however, the spin-valve effect is strongly 
reduced, and conductance is enhanced, except in the regime close to 
antiparallel magnetization, $\phi=\pi$. 
The conductance, then, stays almost flat over a broad range, and then 
establishes the spin-valve effect only in a small region around $\phi=\pi$.

\subsection{First-Order Transport in Nonlinear Response}

A rather complete analysis of first-order transport through quantum-dot spin 
valves, which covers both the linear- and nonlinear-response regime is 
presented in Ref.~\cite{braun}.
There, we derive generalized rate equations for the dot's occupation and 
accumulated spin, which provide the basis of quite an intuitive understanding 
of the behavior of the quantum-dot state.

In the non-linear-response regime, the physics of spin accumulation is more
involved as for linear response.
The accumulated spin tends to align anti-parallel to the drain electrode,
leading to a spin blockade, i.e., a stronger spin-valve effect.
This contrasts with the exchange field which, by rotation of the accumulated
spin, tends to weaken the spin-valve effect.
By the interplay of these two countersteering mechanisms, a very pronounced
negative differential conductance is predicted.

\subsection{Second-Order Transport}

While first-order transport does not probe the spin splitting generated by 
the exchange field, second-order transport does.
Therefore, in second-order transport, the exchange splitting plays a role even 
for collinear configuration of the leads' magnetizations. 
For parallel alignment, the exchange field gives rise to a gate-voltage 
dependent, finite spin polarization of the dot,
$n_{\uparrow} \neq n_{\downarrow}$, even at zero bias. 
This polarization vanishes (at zero bias) for antiparallel orientation
and symmetric coupling, since, in this case, the total exchange field adds up 
to zero. 
A detailed analysis of this transport regime will be presented in 
Ref.~\cite{weymann}, which includes, among other things, the prediction and 
explanation of a peculiar zero-bias behavior for some circumstances.

\subsection{The Kondo Effect}

A very sensitive probe to the exchange field is provided by the
Kondo effect, which occurs in singly-occupied quantum dots below a
characteristic temperature, $k_{\rm B}T \le k_{\rm B}T_{\rm K} \ll \Gamma$. 
The singly-occupied dot defines a local spin with two degenerate states, spin 
up and down. 
The local spin can be flipped by higher-order tunneling processes, in which 
the electron tunnels out of the dot, and another one with opposite spin enters 
from one of the leads. 
By these processes, the dot- and the lead-electron spins are coupled to each 
other. 
At low temperature, a highly-correlated state is formed, in which the local 
spin is totally screened. 
This Kondo-correlated state is accompanied with an increased transmission 
through the dot, and gives rise to a sharp zero-bias anomaly in the 
current-voltage characteristics.

How does a finite spin polarization in the leads modify this picture? 
As it turns out, there are two mechanisms influencing the Kondo effect. 
First, the exchange field lifts the spin degeneracy on the quantum dot. 
This is analogous to the situation of a Kondo dot in the presence of an 
external magnetic field. 
For the latter it is well known, that the zero-bias anomaly splits by twice 
the Zeeman energy. 
Due to the same reason, the exchange-field induces a splitting of the 
zero-bias anomaly for our model system, but now in the absence of an external 
magnetic field. 
In the presence of an external magnetic field both exchange- and magnetic-field
induced splittings contribute. 
In particular, for a properly-tuned magnetic field the level splitting is 
compensated, and a single zero-bias anomaly is recovered.

The second mechanism by which the finite spin-polarization influences the 
Kondo effect is the screening of the quantum-dot spin. 
Naturally, both up- and down-spin electrons in the leads are crucial for the 
screening. 
An imbalance of majority and minority spins in the leads, therefore, weakens 
the screening capability.
As we will see below, this leads to a reduced Kondo temperature $T_{\rm K}(p)$,
which even vanishes for $p=1$.

Recently, the possibility of the Kondo effect in a quantum dot attached
to ferromagnetic electrodes was discussed in a number of publications
\cite{sergueev,zhang,bulka,lopez,martinek1,martinek2,choi}, and it
was shown, that the Kondo resonance is split and suppressed in the
presence of ferromagnetic leads \cite{martinek1,martinek2,choi}. 
It was shown that this splitting can be compensated by an appropriately tuned
external magnetic field to restore the Kondo effect \cite{martinek1,martinek2},
as we discuss in detail below.

In the following, we mainly concentrate on the case of parallel alignment of 
the leads' magnetization. 
For antiparallel alignment and symmetric coupling to the left and right lead, 
the exchange field vanishes (at zero bias voltage), and the usual Kondo 
resonance as for nonmagnetic electrodes forms.

\subsubsection{Perturbative-Scaling Approach.}

An analytical access to the problem, which provides an intuitive picture of 
the involved physics, is the perturbative-scaling approach. 
For detail of the following calculations we refer to Ref.~\cite{martinek1}. 
We make use of the poor man's scaling technique \cite{anderson}, performed in 
two stages \cite{haldane}.
In the first stage, when high-energy degrees of freedom are integrated out, 
charge fluctuations are the dominant. 
Afterwards, in the second stage, we map the resulting model to a Kondo
Hamiltonian, and integrate out the degrees of freedom involving spin 
fluctuations. 
As we will see, each of the two stages will account for one of the two 
above mentioned different mechanisms by which the spin-polarized leads 
influence the Kondo effect, respectively.

The scaling procedure starts at an upper cutoff $D_0$, given by the onsite 
repulsion $U$. 
Charge fluctuations lead to a renormalization of the level position 
$\epsilon_\sigma$ according to the scaling equations

\begin{eqnarray}
  \frac{d \epsilon_\sigma }{d \ln( D_0/D)}
  = |t|^2 \rho_{ \bar{\sigma} } \, ,
\label{eq:Hal_scaling}
\end{eqnarray}

where $\bar \sigma$ is opposite to $\sigma$. 
Since the renormalization is spin dependent, a spin splitting is generated.
In the presence of a magnetic field, this generated spin splitting simply adds
to the initial Zeeman splitting $\Delta \epsilon$.
We obtain the solution $\Delta \widetilde{\epsilon} = \widetilde
\epsilon_\uparrow - \widetilde \epsilon_\downarrow = - (1/\pi) p
\Gamma \ln(D_0/D) + \Delta \epsilon $.
The scaling of (\ref{eq:Hal_scaling}) is terminated \cite{haldane} at 
$\widetilde{D} \sim -\widetilde{\epsilon}$. 
When plugging in $D_0=U$ and $D=\epsilon$, we recover that the generated level 
splitting exactly reflects the zero-temperature limit of the exchange field 
(\ref{exchange}).

To reach the strong-coupling limit, we tune the external magnetic field 
$B_{\rm ext}$  such that the total effective Zeeman splitting vanishes,
$\Delta \widetilde{\epsilon} = 0$. 
In the second stage of Haldane's procedure \cite{haldane}, spin fluctuations 
are integrated out. 
To accomplish this, we perform a Schrieffer-Wolff transformation 
\cite{hewson-book} to map the Anderson model (with renormalized parameters 
$\widetilde{D}$ and $\widetilde{\epsilon}$) to a Kondo Hamiltonian, see
(\ref{spinham}). 
Since we are interested in low-energy excitations only, we neglect the energy 
dependence of the coupling constants and arrive at

\begin{eqnarray}
  H_{\rm Kondo} &=& J_+ S^+ \sum_{rr'kq}a^\dagger_{rk\downarrow}a_{r'q\uparrow}
  + J_- S^- \sum_{rr'kq} a^\dagger_{rq\uparrow}a_{r'k\downarrow}
\nonumber \\
  &&
  + S^z \left( J_{z\uparrow} \sum_{rr'qq'}
  a^\dagger_{rq\uparrow} a_{r'q'\uparrow}
  - J_{z\downarrow} \sum_{rr'kk'}
  a^\dagger_{rk\downarrow}a_{r'k'\downarrow}
  \right)
\, ,
\label{Kondo}
\end{eqnarray}

plus terms independent of either dot spin or lead electron operators,
with $J_+ = J_- = J_{z\uparrow} = J_{z\downarrow} = 
{|t|^2 /|\widetilde \epsilon|} \equiv J_0$ in the large-$U$ limit. 
Although initially identical, the three coupling constants 
$J_+ = J_- \equiv J_\pm$, $J_{z\uparrow}$, and $J_{z\downarrow}$ are 
renormalized differently during the second stage of scaling. 
The scaling equations are 

\begin{eqnarray}
   \frac{ d (\rho_\pm J_\pm) }{d \ln(\widetilde D/D) }
   &=& \rho_\pm J_\pm
   (\rho_\uparrow J_{z \uparrow} + \rho_\downarrow J_{z \downarrow} )
\label{eq:scaling1}
\\
   \frac{ d (\rho_\sigma J_{z\sigma}) }{d \ln(\widetilde D/D)} &=&
   2 (\rho_\pm J_\pm)^2
\label{eq:scaling2}
\end{eqnarray}

with $\rho_\pm = \sqrt{ \rho_\uparrow \rho_\downarrow }$,
$\rho_\sigma \equiv \sum_r \rho_{ r \sigma} $.
To solve these equations we observe that $(\rho_\pm J_\pm)^2 -
(\rho_\uparrow J_{z\uparrow}) (\rho_\downarrow J_{z\downarrow}) =0$ and 
$\rho_\uparrow J_{z\uparrow} - \rho_\downarrow
J_{z\downarrow} = J_0 p (\rho_\uparrow + \rho_\downarrow)$ is constant as well.
I.e., there is only one independent scaling equation. 
All coupling constants reach the stable strong-coupling fixed point 
$J_\pm = J_{z\uparrow} = J_{z\downarrow} = \infty$ at the Kondo energy scale, 
$D \sim k_B T_K$. 
For the parallel configuration, the Kondo temperature in leading order,

\begin{equation}
  T_{\rm K} (P) \approx \widetilde D \exp \left\{ - {1\over
    (\rho_\uparrow + \rho_\downarrow) J_0} \, {{\rm artanh} (p) \over p}
  \right\} \; ,
\label{eq:Kondo_temperature}
\end{equation}

depends on the polarization $p$ in the leads. 
It is maximal for nonmagnetic leads, $p=0$, and vanishes for $p \rightarrow 1$.

The unitary limit for the P configuration can be achieved by tuning the 
magnetic field appropriately, as discussed above. 
In this case, the maximum conductance through the quantum dot is 
$G_{{\rm max},\sigma}^P = e^2/h$ per spin, i.e., the same as for nonmagnetic 
leads.

\subsubsection{Numerical Renormalization Group.}

Although perturbative scaling provides an instructive insight in the 
relevant physical mechanisms, it is a approximate method, and its reliability 
is, a priori, not clear.
The numerical renormalization-group (NRG) technique \cite{hewson-book}, on
the other hand, is one of the most accurate methods available to
study strongly-correlated systems in the Kondo regime.
Recently, it was adapted to the case of a quantum dot coupled to ferromagnetic 
leads \cite{martinek2,choi}.

The NRG study \cite{martinek2,choi} confirms the predictions of the 
perturbative scaling analysis.
The Kondo resonance is split, as a consequence of the exchange field.
By appropriately tuning an external magnetic field, this splitting can be 
fully compensated and the Kondo effect can be restored \cite{martinek2}.
Precisely at this field, the occupancy of the local level is the same for spin 
up and down, 
$\langle n_\uparrow \rangle = \langle n_\downarrow \rangle $, a fact that 
follows from the Friedel sum rule.
Moreover, the Kondo effect has unusual properties such as a strong
spin polarization of the Kondo resonance and for the density of states. 
Nevertheless, the quantum dot conductance is found to be the same for 
each spin channel, $ G_{\uparrow} = G_{\downarrow}$.
Furthermore, by analyzing the spin spectral function, the Kondo temperature
can be determined, and the functional dependence on $p$ as given by 
(\ref{eq:Kondo_temperature}) has been confirmed.

More recently, the NRG scheme has been extended to account for structured 
densities of states \cite{martinek4}.
The generated spin splitting found in this case is found to coincide with
the exchange field defined in (\ref{exchange}), when the energy-dependent
density of states is included in the integral.

\subsubsection{Nonequilibrium Transport Properties.}

To get a qualitative understanding of how the exchange field appears in 
nonlinear transport, we employ an equations-of-motions scheme with the usual 
decoupling scheme \cite{meir}, but generalized by a self-consistent 
determination of the level energy to account for the exchange field in a 
correct way. 
We skip all technical details here (they are given in Ref.~\cite{martinek1}),
and go directly to the discussion of the results.

\begin{figure}
\centerline{\includegraphics[width=10.cm]{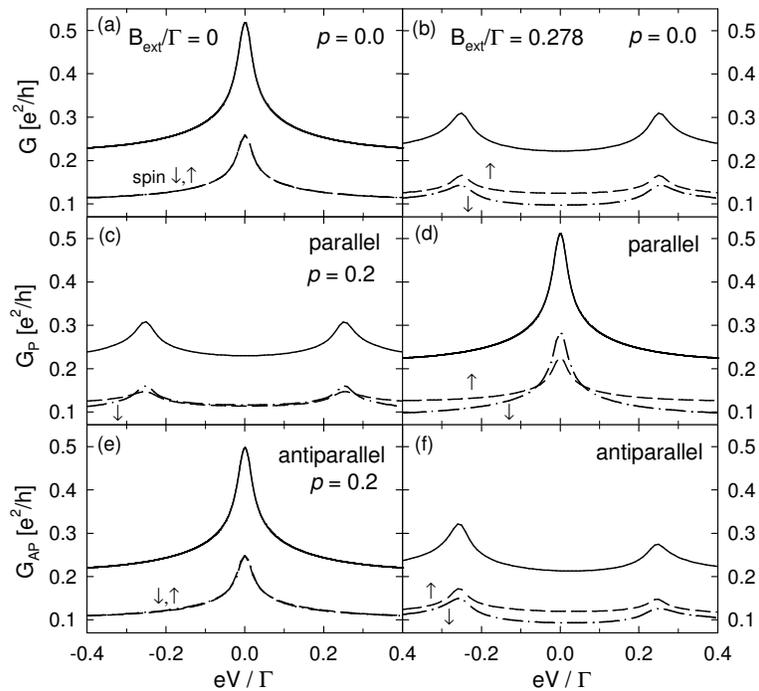}}
\caption{
  Total differential conductance (solid lines) as well as the contributions 
  from the spin up (dashed) and the spin down (dotted-dashed) channel vs. 
  applied bias voltage $V$ at zero magnetic field $ B_{\rm ext} = 0 $ (a,c,e) 
  and at finite magnetic field (b,d,f) for normal (a,b) and ferromagnetic 
  leads with parallel (c,d) and antiparallel (e,f) alignment of the lead 
  magnetizations.  
  The degree of spin polarization of the leads is $ p = 0.2 $ and the other 
  parameters are: $ k_{\rm B}T/\Gamma = 0.005 $ and $ \epsilon /\Gamma = -2 $
}
\label{figure6}
\end{figure}

In Fig.~\ref{figure6} we show the differential conductance as a
function of the transport voltage. 
For nonmagnetic leads, there is a pronounced zero-bias maximum 
[Fig.~\ref{figure6}(a)], which splits in the presence of a magnetic field
[Fig.~\ref{figure6}(b)]. 
For magnetic leads and parallel alignment, we find a splitting of the peak in 
the absence of a magnetic field [Fig.~\ref{figure6}(c)], which can be tuned 
away by an appropriate external magnetic field [Fig.~\ref{figure6}(d)]. 
In the antiparallel configuration, the opposite happens, no splitting at 
$B_{\rm ext}=0$ [Fig.~\ref{figure6}(e)] but finite splitting at 
$B_{\rm ext}>0$ [Fig.~\ref{figure6}(f)] with an additional asymmetry in the 
peak amplitudes as a function of the bias voltage.

We conclude by mentioning that very recent experimental results 
\cite{nygard2,ralph2} indicate confirmation of our theoretical predictions.

\section{Summary}

The interplay of charge and spin degrees of freedom in quantum dots
coupled to ferromagnetic leads is investigated theoretically.
The simultaneous presence of both spin polarization in the leads and
strong Coulomb interaction in the quantum dot generates an exchange field
that acts on the quantum-dot electrons.
We analyze its influence on the dot state and the conductance for different
transport regimes.
Two mechanisms, which can be important, are identified.
The exchange field can precess an accumulated quantum-dot spin, and it 
generates a level splitting.
In the limit of weak dot-lead coupling, the spin precession leads to a 
nontrivial dependence of the linear conductance on the angle between the 
leads' magnetization.
For strong dot-lead coupling, the exchange field is detectable in a 
splitting of the Kondo resonance, which can be tuned away by additionally 
applying an external magnetic field.

\section{Acknowledgments}

The presented work is based on joint publications with 
L. Borda, M. Braun, R. Bulla, J. von Delft, H. Imamura, S. Maekawa, M. Sindel, 
Y. Utsumi, and I. Weymann, all of whom we thank for fruitful collaboration.

We thank G. Bauer, A. Brataas, P. Bruno, T. Costi, A. Fert, L. Glazman, 
W. Hofstetter, B. Jones, C. Marcus, J. Nyg{\aa}rd, A. Pasupathy, 
D. Ralph, A. Rosch, S. Takahashi, D. Urban, and M. Vojta for discussions. 
This work was supported by the Deutsche Forschungsgemeinschaft under
the Center for Functional Nanostructures and the Emmy-Noether program, and
by the European Community under 
the 'Spintronics' RT Network of the EC RTN2-2001-00440, 
Project PBZ/KBN/044/P03/2001,
and the Centre of Excellence for Magnetic and
Molecular Materials for Future Electronics within the EC Contract
G5MA-CT-2002-04049.


\begin{thebibliography}{8.}
\bibitem{com_models}
We note that, for all the effects discussed in this paper, the 
(spin-dependent) density of states $\rho_\sigma$ and the (spin-independent) 
tunneling amplitudes $t$ enter via the combination $\rho_\sigma |t|^2$. 
As a consequence, all the conclusions drawn in this paper are valid also for 
a system with spin-dependent tunneling amplitudes but nonmagnetic leads, or
a mixture of both.
\bibitem{comment_2}
In equilibrium and zeroth order in $\Gamma$, the average spin on the dot
vanishes: majority-spin electrons enter the quantum dot at a higher rate than 
minority spins, but this is exactly compensated by a higher rate for the 
majority spins to leave the dot.
\end{thebibliography}
\end{document}